# Fragility of Unidirectional Transport in Weakly Disordered Photonic Chern Insulators

Xiaoxuan Shi [1†], Tiantao Qu [1†], Xianbin Wu [1], Mudi Wang [2], Lei Zhang [3,4*] and Jun Chen [1,4*]

[1]State Key Laboratory of Quantum Optics Technologies and Devices, Institute of Theoretical Physics, Shanxi University, Taiyuan 030006, China

[2]Key Laboratory of Artificial Micro- and Nanostructures of Ministry of Education and School of Physics and Technology, Wuhan University, Wuhan 430072, China.

[3]State Key Laboratory of Quantum Optics Technologies and Devices, Institute of Laser Spectroscopy, Shanxi University, Taiyuan 030006, China

[4]Collaborative Innovation Center of Extreme Optics, Shanxi University, Taiyuan 030006, China

*Correspondence to: zhanglei@sxu.edu.cn; chenjun@sxu.edu.cn

**ABSTRACT:** Photonic Chern insulators enable unidirectional light transport protected by nontrivial band topology—essential for robust photonic integrated circuits and error-free communication. However, disorder from impurities or defects inevitably exists in practical applications, yet how weak disorder affects topological chiral edge states remains insufficiently understood. Here, we reveal a previously unrecognized mechanism by which weak disorder can disrupt robust propagation of chiral edge states in photonic Chern insulators, despite the preservation of global topological invariants. By randomly replacing a small number of magnetized rods with nonmagnetized impurities in a magnetic photonic crystal, we find that when the excitation frequency approaches the single impurity defect state frequency, weak coupling between spatially extended defect states forms a topologically trivial impurity band inside the topological gap. This enables coexistence and coupling of defect states and chiral edge states. The reciprocal "necklace state" transport channels formed by coupled defect states break the expected unidirectional propagation in topological Chern insulators with weak disorder. Our work reveals that topological chiral edge state and disorder interactions are more intricate than previously understood and provides new insights into stability and control of topological transport in realistic applications.

## Introduction

Topological photonic systems[1–9] provide an effective platform for exploring robust wave transport immune to impurities and defects. Analogous to quantum Hall systems in electronics,[10–12] photonic Chern insulators with broken time-reversal symmetry host topological chiral edge states that enable unidirectional, backscattering-immune light propagation along their boundarie.[13–16] These chiral edge states arise from the nontrivial topology of bulk band structures characterized by quantized Chern numbers.[15,17–19] Their unidirectional propagation is therefore generally considered highly robust, and has been realized in systems such as gyromagnetic photonic crystals[20–25] and Floquet topological insulators.[26–31] Due to their immunity to local perturbations and structural defects, chiral edge states are ideal candidates for robust topological waveguides,[32–36] topological lasing,[37–41] and on-chip photonic communication.[42–46]

However, disorder[47] arising from randomly distributed impurities or intrinsic structural defects is inevitable in real materials and practical applications, making research on disorder-induced effects essential. Research on disordered systems has predominantly focused on disorder regimes[48–50] that can close or reopen the mobility gap,[49–51] fundamentally altering the system's topological properties. For instance, in amorphous systems, enhanced structural disorder can close the bulk mobility gap, triggering a topological phase transition from a glass-like to a liquid-like lattice and leading to the disappearance of topological edge states.[49,52,53] Conversely, introducing moderate structural disorder or on-site disorder potentials into an initially topologically trivial system can induce a gap closing and reopening, driving a transition from a trivial insulator to a topological Anderson insulator.[50,51,54–59] In studies investigating the evolution of topological properties across varying disorder strengths, it is generally accepted that weak disorder has a negligible impact on established phases, allowing systems to robustly maintain their protected edge states and quantized transport.[49–62] Consequently, localized weak disorder—such as edge defects including missing unit cells or obstacles—is frequently used to verify the robustness of unidirectional chiral edge state propagation.[16,22,55] This raises a critical question: does weak disorder truly leaves the unidirectional propagation of chiral edge states intact?

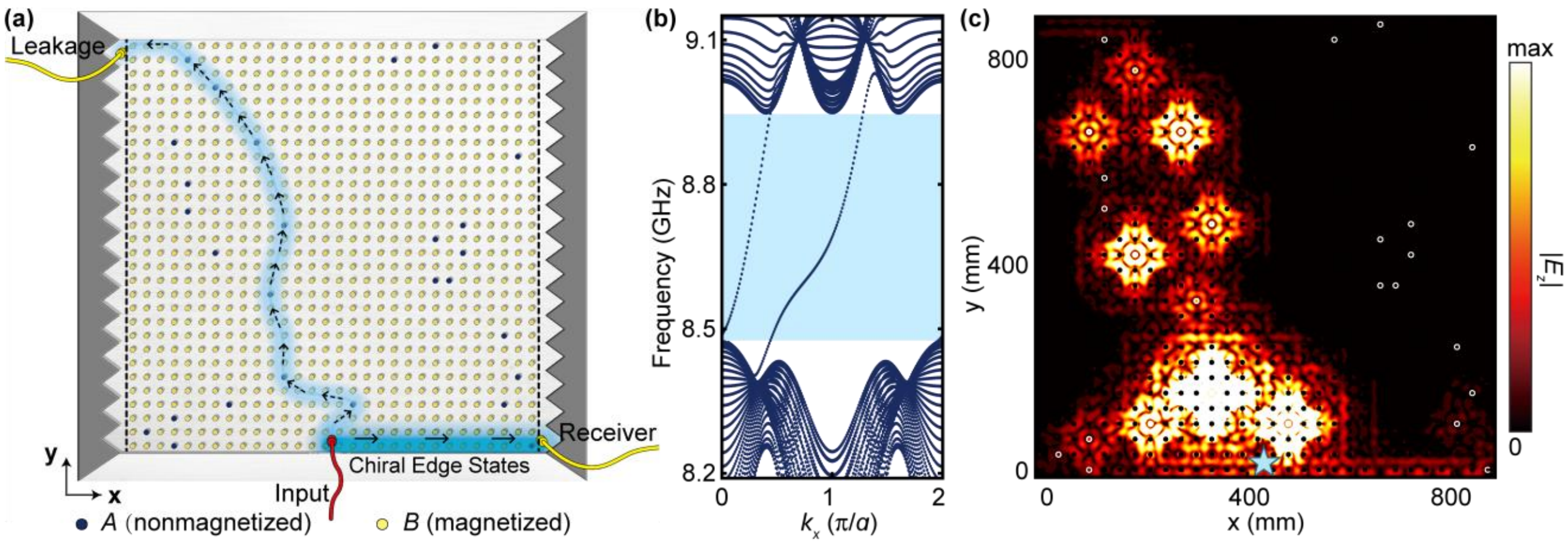

Figure 1. Disruption of the robust unidirectional propagation of chiral edge states by weak disorder. (a) Schematic diagram showing the influence of weak disorder on optical paths in a magnetic square lattice photonic crystal with impurities [created by randomly replacing some magnetized gyromagnetic rods ($B$-type, yellow dots) with nonmagnetized rods ($A$-type, dark blue dots)]. Additional details on materials and configuration are provided in Methods. Absorbing boundary conditions are applied to the left and right boundaries. (b) TM mode dispersion curves for the pure magnetic photonic crystal, showing a topological band gap with Chern number 2. The two dotted lines in the blue band gap region represent the two chiral edge modes. (c) Electric field distribution $|E_z|$ under line source excitation with 3% impurity concentration. The excitation frequency is 8.836 GHz. Hollow white circles represent nonmagnetized rods (impurities), black dots represent magnetized rods, and the blue star marks the line source location.

In this work, we demonstrate that weak disorder (specifically low-concentration impurity doping) can disrupt the robust propagation of chiral edge states in photonic Chern insulators. By introducing nonmagnetized impurities into a magnetized photonic crystal composed of gyromagnetic rods, we show that weak coupling between impurity-induced defect states can form reciprocal "necklace state" transport channels. The most effective coupling occurs when the excitation frequency approaches the single impurity defect state frequency. This coupling creates a topologically trivial impurity band around that frequency. Since this trivial band lies between two nontrivial topological gaps—both containing chiral edge states—the chiral edge states "pass through" the impurity band and coexist with the defect states. This leads to the coupling between "necklace state" and chiral edge states, disrupting the unidirectional propagation robustness of chiral edge states. Importantly, introducing weak disorder does not close the topological gap or alter the global topological invariants, even when unidirectional propagation breaks down. The degradation of robust chiral edge propagation arises from resonant scattering driven by "necklace states" that form a competing leakage pathway, rather than from a fundamental breakdown of the bulk topology. Furthermore, while "necklace states" have been extensively investigated in the context of standard Anderson localization[63–71]—where multi-resonance tunneling dominates transport in strongly localized 1D or 2D trivial systems—their specific coexistence and interaction with topologically protected edge states has remained unexplored. Our findings uncover a previously unrecognized interplay between these spatially extended defect states and topological chiral edge states, demonstrating that weak disorder induced trivial defect modes can disrupt topological transport even when the global topological phase remains intact.

## Results

***Disruption of the robustness of chiral edge state propagation by weak disorder***

Let's start with an ordinary two-dimensional square lattice photonic crystal composed of gyromagnetic rods (Figure 1a). Each rod can be magnetized individually by attaching permanent magnets to its ends, applying a static magnetic field along the $z$ direction.[72] When all gyromagnetic rods are magnetized, this magnetic photonic crystal exhibits a topological band gap with a Chern number of 2, ranging from 8.476 to 8.946 GHz in the transverse magnetic (TM) band structures (Figure 1b). This gap is accompanied by two unidirectionally propagating edge modes. It is generally believed that these chiral edge states are topologically protected, making their unidirectional propagation highly robust against weak disorder caused by impurities.[15,16,53,55] This robust characteristic of unidirectional light ("information") transport has inspired practical applications in various fields, especially in error-free communication through topological chip designs.[42,43]

However, we surprisingly find that weak disorder caused by impurities may seriously damage the robustness of the chiral edge states' unidirectional propagation (see Figure 1c). Specifically, the disorder is created by randomly removing permanent magnets from the ends of some magnetized rods in the magnetic photonic crystal. We define the impurity concentration $x_c$ as $x_c = N_A/(N_A + N_B)$, where $N_A$ and $N_B$ represent the number of nonmagnetized and magnetized gyromagnetic rods, respectively. Even within a random configuration with the impurity concentration of only 3% in Figure 1c, the robustness of the chiral edge states' unidirectional propagation is disrupted. The coupling between impurity-induced defect states,[74–76] as well as between defect states and chiral edge states, creates new optical paths. This coupling can even excite chiral edge states on the opposite boundary, leading to "information leakage" (see also the schematic diagram of optical paths in Figure 1a). It should be noted that this phenomenon does not only affect chiral edge states in high Chern number band gaps. We will later show that weak disorder may also disrupt the robust propagation of the chiral edge state in a band gap with a Chern number of −1.

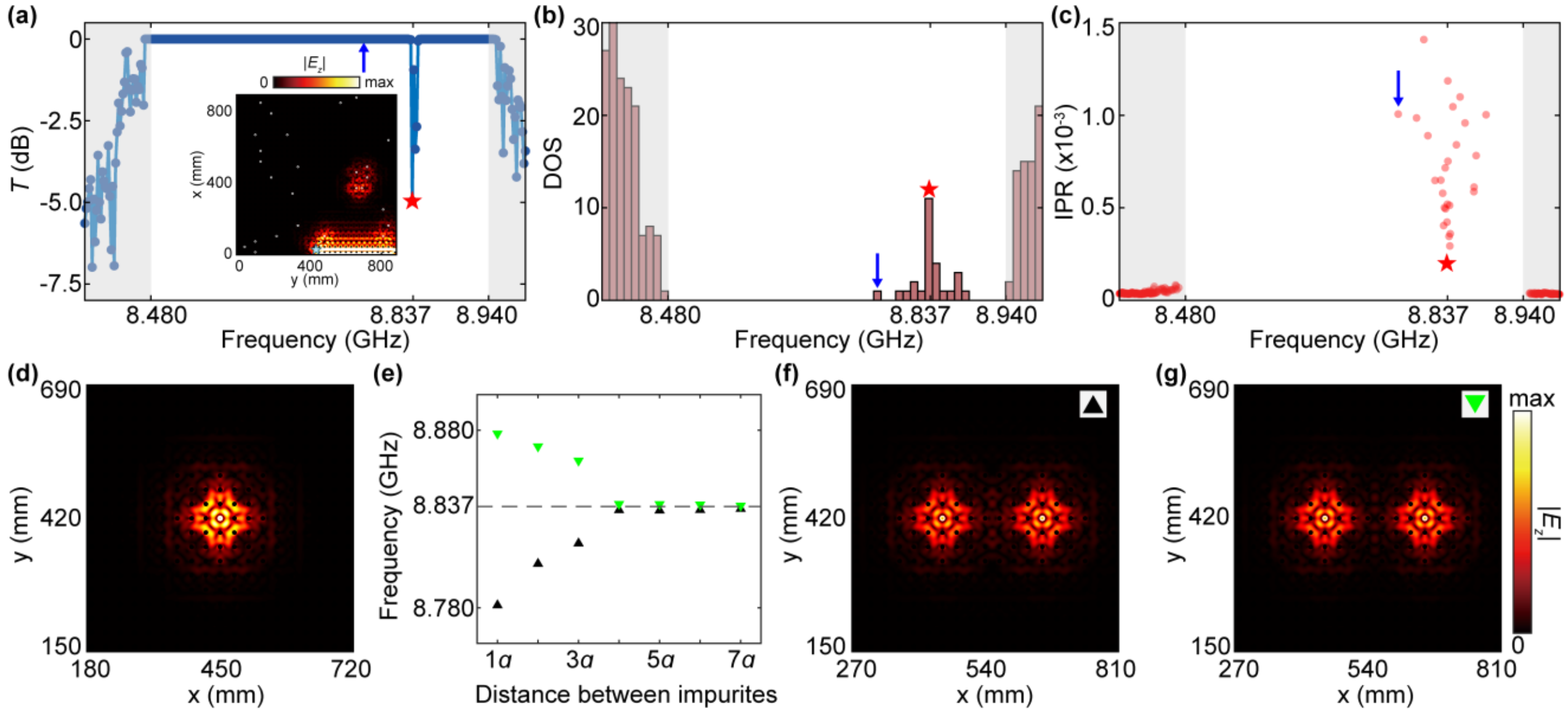

**Figure 2.** Analysis of factors affecting unidirectional propagation robustness of chiral edge states under weak disorder. (a) Unidirectional transport efficiency of chiral edge states at different excitation frequencies for the configuration shown in Figure 1a (3% impurity concentration). Energy passing through the system via the left and right boundaries is calculated by integrating the time-averaged Poynting vector over the left and right black dashed lines in Figure 1a. The red star corresponds to the situation in Figure 1c. The inset shows the electric field distribution at excitation frequency 8.770 GHz, indicated by the blue arrow. (b) Simulated DOS showing the number of eigenstates excited at different frequencies for the same configuration. (c) IPR of the corresponding states in (b), describing the localization properties of the states. (d) Electric field distribution of the defect state when only one nonmagnetized rod impurity exists in the system. (e) Defect state coupling frequencies for the bonding (black triangles) and anti-bonding modes (green triangles) versus distance between the two impurities. The horizontal dashed line at 8.837GHz indicates the single-impurity defect state frequency. (f, g) Electric field distributions of the (f) bonding and (g) antibonding modes resulting from defect state coupling when only two impurities (separated by $7a$) exist in the system. Other parameters in (d−g) are the same as in Figure 1a. Light gray shaded regions in (a—c) represent bulk regions. Hollow white circles in the inset of (a) and in (d, f,g) represent nonmagnetized rods (impurities), and black dots represent magnetized rods.

In fact, weak disorder disrupting the robustness of chiral edge state propagation is indeed a rare phenomenon. If we define the unidirectional transport efficiency of chiral edge states as

$$T = 20\log_{10}\left(\frac{E_{\text{out}}}{E_{\text{tot}}}\right), \tag{1}$$

where $E_{\text{out}}$ is the energy passing through the system (Figure 1a) via the right boundary, and $E_{\text{tot}}$ is the total energy leaving the system from both the left and right boundaries. We find that, for the configuration shown in Figure 1a, weak disorder disrupts the robustness of the chiral edge states' unidirectional propagation only at specific frequencies (around 8.837 GHz), causing a sharp drop in unidirectional transport efficiency $T$ (Figure 2a). The red star marks the excitation frequency that produces the situation in Figure 1c. At most other frequencies, chiral edge states maintain robust transport ($T = 0$) even with impurities present, as shown in the inset of Figure 2a (blue arrow indicates excitation frequency). This is typical behavior in weakly disordered topological systems. Comparing the field distributions in Figure 1c and the inset of Figure 2a reveals that weak disorder disrupts unidirectional propagation robustness of chiral edge states when defect states couple effectively. For effective coupling to occur with few impurities, two conditions should be met: exciting as many defect states as possible simultaneously, and ensuring these defect states are sufficiently extended to couple with each other. Figures 2b,c illustrate this well. The density of states (DOS) in Figure 2b is calculated by treating the whole square array as a supercell with periodic boundary conditions. The distribution of defect states in the gap shows that at approximately 8.837 GHz, many defect states are excited simultaneously. The excitation frequency where weak disorder disrupts the robust propagation in Figure 1c falls at this peak (red star). In contrast, the excitation frequency for the inset of Figure 2a corresponds to the blue arrow in Figure 2b. Despite having the same number of impurities, few defect states are excited at this frequency, as also seen in the field distribution in the inset of Figure 2a. Figure 2c further shows the localization behavior of these defect states at different eigenfrequencies through the inverse participation ratio (IPR).[77,78] The IPR for each state in the photonic system is calculated by[72]

$$\text{IPR}(\omega) = \frac{\sum_i^N \sum_j^M \left|E_z(\omega, x_i, y_j)\right|^4}{\left(\sum_i^N \sum_j^M \left|E_z(\omega, x_i, y_j)\right|^2\right)^2}, \tag{2}$$

where $\omega$ is the working frequency, $E_z$ is the electric field of each state at coordinates $(x_i, y_j)$, $N$ and $M$ are the number of coordinates calculated, and the sum runs over all points in real space. If a state is highly localized—confined to only one coordinate point—then its IPR = 1. Conversely, the more extended a state is, the smaller its IPR value. For example, bulk states in the gray shaded region have IPR values approaching 0. Figure 2c shows that around 8.837 GHz, many defect states (each red dot represents an eigenstate) are extended with small IPR values. Combined with Figure 2b, this explains the situation in Figure 1c (marked by the red star): weak disorder disrupts propagation robustness of chiral edge states around 8.837 GHz because many de fect states are excited simultaneously, and a considerable portion of these states are sufficiently extended to enable effective coupling between defect states and with chiral edge states. In contrast, at most other frequencies—such as the one indicated by the blue arrow, with the field distribution shown in the inset of Figure 2a—fewer defect states are excited, and these states are more localized (with relatively large IPR values). As a result, effective coupling between defect states cannot occur to generate new optical paths. We have confirmed this conclusion through statistical analysis of more random configurations (see Supporting Information, section S1).

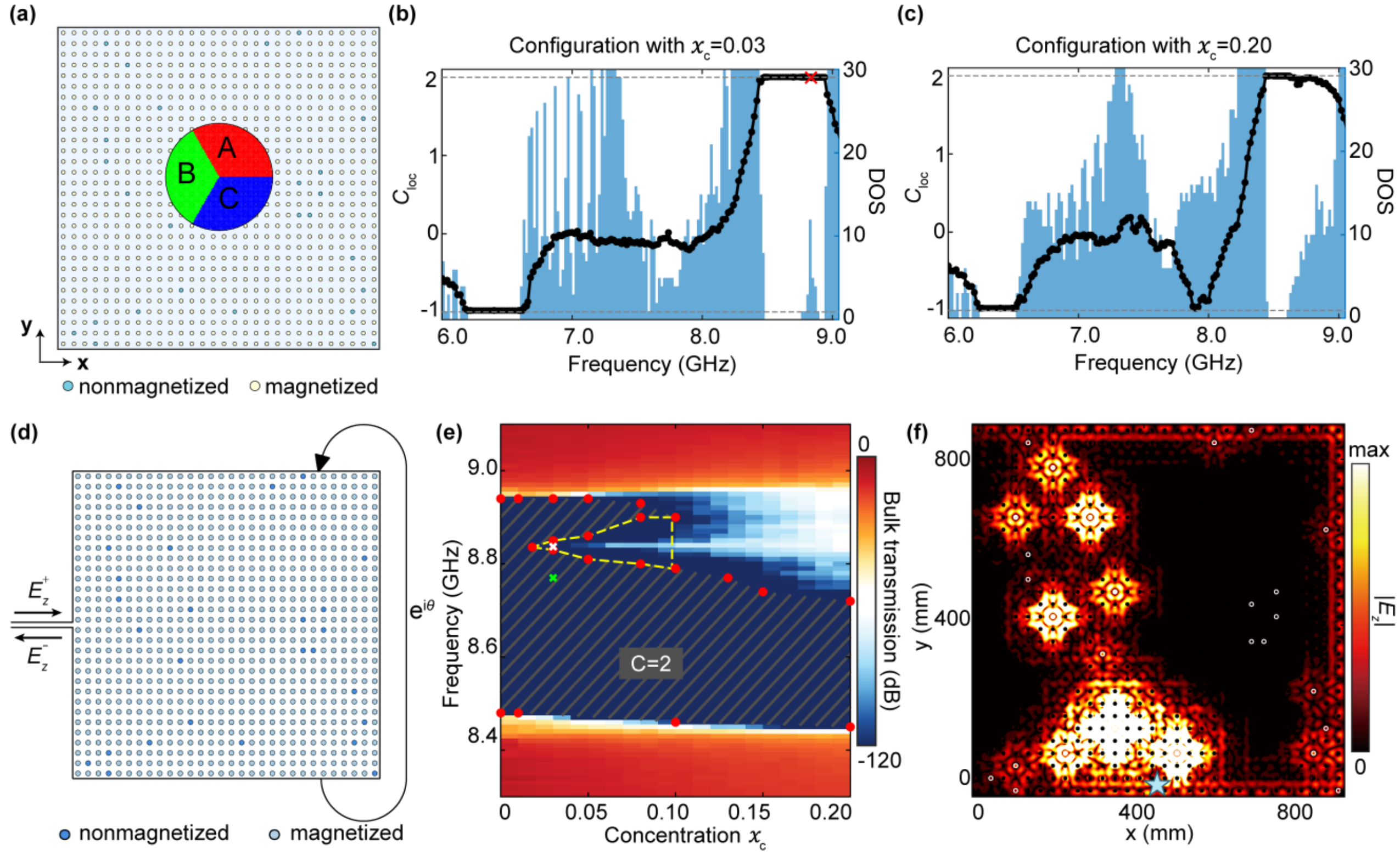

**Figure 3.** Study of topological properties of weakly disordered photonic systems. (a) Schematic diagram for calculating the local Chern number (LCN) $C_{\mathrm{loc}}$ of a magnetic square lattice photonic crystal with impurities. The summation region is a disk of radius $5a$, discretized into grids with step size $dx = dy = a/20$. The disk is partitioned into three regions (red, green and blue), each containing 10472 grid points. The entire doped system is treated as a supercell surrounded by periodic boundaries. LCN profiles are shown for (b) the configuration used in Figure 1a with impurity concentration $x_{\mathrm{c}} = 0.03$, and (c) a random configuration with impurity concentration $x_{\mathrm{c}} = 0.20$. The red cross in (b) indicates the frequency 8.836 GHz, where the robust propagation of chiral edge states is disrupted. For reference, the blue shaded histograms in (b) and (c) illustrate the corresponding statistical distributions of DOS versus frequency. (d) Schematic diagram of reflection phase winding calculation. (e) Bulk transmission spectrum as a function of impurity concentration $x_{\mathrm{c}}$. Each data point represents the average of 60 random configurations. The shaded region marks the topological gap with a Chern number of 2. Red dots mark the topological gap boundary. For each impurity concentration, the topological gap is determined using the reflection phase winding method by analyzing the 60 random configurations. (f) Electric field distribution $|E_z|$ under line source excitation for the photonic system with 3% impurity concentration. The configuration used has the same parameters as Figure 1c but without the absorbing boundary condition on the right. Hollow white circles represent nonmagnetized rod impurities, black dots represent magnetized rods, and the blue star marks the line source location.

Then, another question arises: why this specific frequency of 8.837 GHz? In fact, the most effective coupling occurs when the excitation frequency matches the defect state frequency of a single impurity—which is 8.837 GHz. This can be understood as follows: when only one nonmagnetized rod creates an impurity, its defect state locates at 8.837 GHz (determined by lattice parameters and materials), with its eigenmode field distribution shown in Figure 2d. When multiple impurities are present, coupling between defect states splits the original eigenfrequency, spreading around 8.837 GHz—bonding modes redshift while antibonding modes blueshift (see Figure 2e). Different coupling modes (influenced by the number of impurities, their spacing, and positions) have different frequencies. How can multiple defect modes be excited simultaneously at the same frequency in a weakly disordered system? The answer lies in weak coupling. Weak coupling causes only small frequency shifts, keeping the coupled modes' frequencies nearly identical—close to the single-impurity resonance frequency. As shown by the black and green triangles in Figure 2e and eigenmode field distributions in Figures 2f,g, bonding and antibonding modes from two impurities separated by $7a$ have nearly identical frequencies (8.836 GHz and 8.838 GHz), very close to the single-impurity frequency. Thus, the formation of the "necklace state" transport channel in Figure 1c is attributed to some disordered impurities being separated by $4a$ to $7a$. This spatial configuration facilitates weak coupling among multiple defect states. Because the coupling is weak, the resulting eigenfrequencies remain closely clustered around the single-impurity frequency of 8.837 GHz, as indicated by the frequency distribution in Figure 2e and the eigenstates distribution marked by the red star in Figure 2b. Consequently, weak disorder disrupts the robustness of chiral edge state propagation only when impurities are optimally spaced—neither too close nor too far—enabling weak coupling to generate new optical paths. It follows that if a specific configuration lacks impurities arranged at this optimal spacing, the "necklace state" transport channel will not manifest, regardless of the excitation frequency. In disordered systems, this particular impurity distribution is rare, making robustness breakdown in unidirectional propagation uncommon. Of the 60 random configurations calculated (3% doping), only 2 exhibited this phenomenon (see Figure S2b of Supporting Information). Although this phenomenon is rare under random weak disorder, it is important to note that photonic Chern insulators used for error-free communication must be protected against information leakage and disruption caused by such intentional design flaws.

***Topological properties in weakly disordered systems***

A pressing question remains: when weak disorder disrupts unidirectional propagation robustness of chiral edge states, is the photonic system still topological? The topological properties of periodic photonic systems can be determined by calculating the Berry curvature in $k$-space to obtain the Chern number.[16,19,24] However, photonic crystals with disordered impurities are nonperiodic systems. We can use the local Chern number, calculated via the Kitaev real-space formula,[57,79,80] to evaluate the systems' topological features. As illustrated in Figure 3a, this method utilizes real-space projection operators over local regions A, B, and C, with further technical details provided in section S3 of the Supporting Information. In Figure 3b, we compute the local Chern number for the disordered configuration (3% impurity concentration) used in Figure 1a. The result demonstrates that even at the excitation frequency (8.836 GHz, marked by the red cross) where "necklace states" form and unidirectional transport degrades, the local Chern number remains exactly 2. This rigorous real-space calculation definitively proves that the global nontrivial topology is not destroyed by the weak disorder. As the impurity concentration increases, Figure 3c shows a substantial narrowing of the topological gap at a 20% impurity concentration.

To further understand that the weak disorder induced "necklace state" does not affect the bulk topology, we also employ the reflection phase winding method.[57,72,73] Specifically, as shown in Figure 3d, the left side of the photonic system is bounded by a perfect magnetic conductor and connects to an air waveguide lead supporting a single TM mode at the relevant frequency range. The right side has an absorbing boundary condition. A twisted boundary condition $\Psi(y=L)=\Psi(y=0)e^{i\theta}$ is imposed on the top ($y=L$) and bottom ($y=0$) boundaries. Physically, this twisted boundary condition can be viewed as an adiabatically changing gauge flux $\theta$ threading through the hollow of the rolled-up photonic system.[57,73] The Chern number then equals the winding number of the reflection phase $\varphi(\theta)$ as the twisting angle $\theta$ varies from 0 to $2\pi$. This method requires that the incident waves $E_z^+(\theta,\omega)$ are totally reflected. Only then does the reflection matrix satisfy $R(\theta)=E_z^-(\theta)/E_z^+(\theta)=e^{i\varphi(\theta)}\in U(1)$. When we adiabatically vary $\theta$ over a cycle, the first homotopy group of $U(1)$ equals $\mathbb{Z}$, determining the topological classification of the evolutions of $R(\theta)$.[57] We mark the topological gap with Chern number 2 in the dark blue gap region in Figure 3e with shading. The Chern number of 2 corresponds to two winding cycles, where the reflection phase changes cumulatively by $4\pi$ (see Figure S4a of Supporting Information). A comparison of Figure 3e with Figures 3b and 3c reveals that both methods yield consistent results in characterizing system's topological features such as the gap and invariants. The situation shown in the inset of Figure 2a corresponds to the green cross in Figure 3e. Although impurities exist in the photonic system, the defect states do not affect the bulk transmission, and the chiral edge state propagation remains topologically protected. However, under weak disorder, there exists a region where the reflection phase winding method fails—the region enclosed by the yellow dashed line. For example, Figure 3f corresponds to the white cross in this region. Here, defect states couple to form "necklace state" transport channels, affecting the bulk transmission. The incident waves are no longer totally reflected, causing the reflection phase winding method to fail.

Nevertheless, we can indirectly determine the topological properties of this region. Since the regions both above and below the enclosed region are topologically nontrivial with Chern number 2 (see Supporting Information, section S4), the enclosed region can only possess trivial topological properties. This region can be considered as a topologically trivial impurity band formed by effective coupling of defect states induced by nonmagnetized rod impurities around 8.837 GHz. These coupled states create reciprocal propagation paths for light. Since this topologically trivial impurity band lies between two topologically nontrivial gaps—both containing chiral edge states with Chern number 2—the chiral edge states "pass through" the impurity band and couple with the defect states. As shown in Figure 3f, this creates a coexistence of "necklace state" formed by coupled defect states and chiral edge states, disrupting unidirectional propagation robustness of chiral edge states. As impurity concentration increases, this impurity band in Figure 3e will eventually merge into the bulk region, leading to a concomitant narrowing (see also Figure 3c) or eventual closure of the topological gap.

Thus, it can be concluded that the topological gap does not need to close for unidirectional transport to be compromised. We prove that in the weak disorder regime—defined here as a low-concentration impurity doping regime where the disorder strength is strictly insufficient to close the bulk mobility gap or alter the global topological invariants—optimally spaced defects can resonantly couple to form a leaky channel. This trivial "necklace state" channel perfectly coexists within the topological gap, providing a parallel pathway that circumvents the transport isolation of the chiral edge states. Consequently, in photonic Chern insulators, the emergence of this "necklace state" introduces a competing leakage pathway that disrupts the propagation robustness of chiral edge states, despite the preservation of global topological invariants. Regarding the formation of necklace states, as analyzed above, these sparse impurities positioned at an optimal spacing—approximately twice the localization length of a single impurity—facilitate weak coupling. Beyond this distance, the defect states remain highly localized and isolated. This weak coupling induces only minimal frequency shifts, enabling the simultaneous excitation of multiple defect coupling modes at the same frequency nearly identical to that of a single-impurity defect state. Consequently, the probability of observing a coexistence of "necklace state" and chiral edge states—which leads to unidirectional propagation disruption—peaks at frequencies near the single-impurity defect resonance (see Figure S2 of Supporting Information). Beyond the excitation frequency, the selection of system geometry and materials is crucial to optimize the spatial decay length of the single-impurity defect mode, ensuring defect states are sufficiently extended to facilitate mutual coupling. Consequently, systems with a longer localization length of the single-impurity defect mode are more susceptible to the coexistence of "necklace state" and chiral edge states (see Figure S2 of Supporting Information). However, it must be noted that an excessively long localization length may lead to the closure of the topological gap. Moreover, for the threshold concentration $x_{\mathrm{th}}$ required to observe this disruption, we find that it scales inversely with the system ($Na \times Na$) size: $x_{\mathrm{th}} \sim 1/N$ (see Supporting Information, section S5). As the system size increases, the impurity concentration required to trigger this robust transport disruption becomes progressively smaller. Nevertheless, under fully random disorder sampling, the probability of observing "necklace state" transport channels that break the unidirectional propagation of chiral edge states remains very low at ultralow impurity concentrations. Therefore, we have utilized a 3% concentration here to clearly demonstrate these effects.

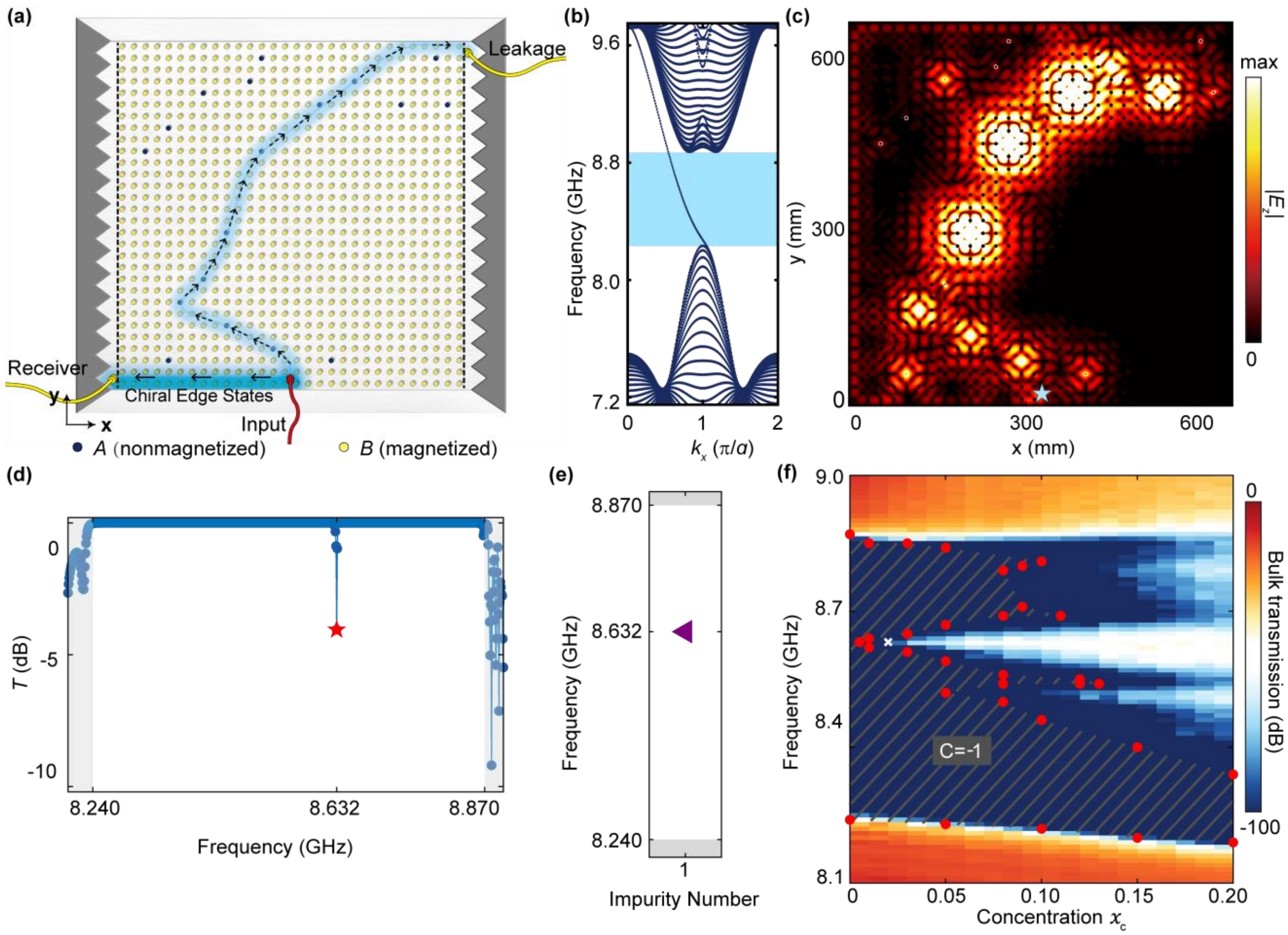

**Figure 4.** Weak disorder disrupting the robust unidirectional propagation of a chiral edge state with Chern number −1. (a) Schematic diagram showing the influence of weak disorder on optical paths in a magnetic square lattice photonic crystal with 2% impurity concentration. This square array ($30a \times 30a$) has a lattice constant $a = 22.5$ mm, magnetized rods with radius $r_B = 0.13a$, and nonmagnetized rod impurities with radius $r_A = 1.15r_B$. Absorbing boundary conditions are applied to the left and right boundaries. (b) TM mode dispersion curves for the magnetic photonic crystal in (a) without impurities, showing a topological band gap with Chern number −1 from 8.234 to 8.874 GHz. The dotted line in the blue band gap region represents the single chiral edge mode. (c) Electric field distribution $|E_z|$ under line source excitation at 8.632 GHz for the configuration in (a), but with the left absorbing boundary removed. Hollow white circles represent nonmagnetized rod impurities, black dots represent magnetized rods, and the blue star marks the line source location. (d) Unidirectional transport efficiency of the chiral edge state at different excitation frequencies. (e) The purple triangle marks the eigenfrequency of 8.632 GHz for the defect state when only one nonmagnetized rod impurity exists in the system. Light gray shaded regions in (d) and (e) correspond to the bulk regions. (f) Bulk transmission spectrum as a function of impurity concentration $x_c$. Each data point represents the average of 60 random configurations. The shaded region marks the topological gap with Chern number −1. Red dots mark the topological gap boundary. For each impurity concentration, the topological gap is determined using the reflection phase winding method by analyzing the 60 random configurations. Panel (c) corresponds to the white cross in this figure.

***Generality across Chern-number regimes***

Finally, it should be noted that weak disorder can disrupt the propagation robustness of chiral edge states with both high and low Chern numbers through similar mechanisms. Consider the magnetic photonic system in Figure 4a. Without nonmagnetized rod impurities, Figure 4b displays the TM mode dispersion curves of this magnetic photonic crystal, showing a chiral edge mode in the topological band gap with Chern number −1 from 8.234 to 8.874 GHz. Similar to the Chern number 2 case analyzed above, when magnetized rods are randomly replaced by nonmagnetized rod impurities—even at only 2% impurity concentration—weak disorder may disrupt the robustness of chiral edge state propagation, as shown in Figure 4c. Impurity-induced defect states effectively couple to form "necklace state" reciprocal transport channels. The emergence of these new optical paths disrupts the robustness of chiral edge state propagation, leading to "information leakage" (see also the optical path schematic in Figure 4a). For Chern number −1, this phenomenon remains rare under random weak disorder. As shown in Figure 4d, the unidirectional transport efficiency (defined as in eq 1, but with $E_{\mathrm{out}}$ redefined as the energy passing through the system via the left boundary) exhibits a sharp drop only around 8.632 GHz—the single-impurity defect state eigenfrequency for the parameters used in this system (see Figure 4e)—corresponding to weak disorder severely affecting the robustness of chiral edge state propagation. The red star corresponds to the situation in Figure 4c. Moreover, among the 60 randomly generated disorder configurations with 2% impurity concentration, only 4 exhibited this disrupted propagation robustness (see Figure S2e of Supporting Information). For Chern number −1, the underlying principle mirrors that for Chern number 2. As shown in Figure 4f, effective coupling of defect states forms a topologically trivial impurity band around the single-impurity defect state frequency 8.632 GHz. Since this topologically trivial impurity band lies between two topologically nontrivial gaps—both containing a chiral edge state with Chern number −1 (see Supporting Information, section S4)—the chiral edge state "passes through" the impurity band and couples with the defect states. The coexistence of "necklace state" formed by coupled defect states and chiral edge states disrupts the unidirectional propagation robustness of the chiral edge state. Additionally, when defect states have stronger spatial extension, more than one topologically trivial impurity band may appear. For example, in Figure 4f, besides the impurity band around 8.632 GHz, additional impurity bands appear around 8.520 GHz and 8.790 GHz. This occurs because the single-impurity defect state in the magnetic photonic system shown in Figure 4a has stronger spatial extension (visible when comparing the electric field distributions in Figure 4c and Figure 2d), which enables stronger coupling between defect states. This stronger coupling causes significant eigenfrequency splitting around 8.632 GHz—bonding modes redshift while antibonding modes blueshift—resulting in multiple impurity bands. Under the parameters used in Figure 4a, the impurity band generated by antibonding modes lies close to the bulk region, making it less likely to exhibit the phenomenon of coupled defect states and chiral edge state coexistence that disrupts the propagation robustness of chiral edge states, compared to the two lower-frequency impurity bands that penetrate deeper into the topological gap.

**Conclusion and discussion**

We have identified that the emergence of "necklace state" undermines chiral edge state unidirectionality even at weak disorder levels. When the excitation frequency approaches the defect state resonance of a single impurity, even a few impurities can couple weakly to form reciprocal "necklace state" channels. These coupled defect states aggregate into a narrow, topologically trivial impurity band that lies within the topological gap. Chiral edge states traversing this frequency range necessarily couple with the impurity band, resulting in the breakdown of unidirectionality. When the localization length (i.e., the spatial decay length) of a single-impurity defect mode is large, the enhanced spatial extension facilitates stronger coupling between impurities. This strong interaction leads to significant eigenfrequency splitting around the single-impurity resonance, resulting in the formation of multiple impurity bands. Consequently, "necklace state" may emerge at frequencies detuned from the single-impurity frequency. However, an excessively long localization length will eventually lead to the closure of the topological gap. We emphasize that weak disorder in this context refers to a low-concentration impurity doping regime where the disorder strength is strictly insufficient to close the bulk mobility gap or alter the global topological invariants.

This breakdown does not contradict topological protection but reveals a subtle, previously overlooked interaction between edge modes and spatially extended defect states. The spatial extension and near-degeneracy of weakly coupled defect modes allow them to collectively influence the robustness of chiral edge state propagation, despite the disorder being far below the threshold typically associated with bulk topology changes. Because this mechanism requires both resonant excitation and optimal impurity spacing, its occurrence in random configurations is statistically rare. However, because the threshold concentration needed to trigger this breakdown scales inversely with system size, even rare breakdown pathways become technologically significant for large-scale photonic devices with stringent performance requirements—not to mention information leakage and disruption caused by such intentional design flaws.

Our findings demonstrate that the interplay between topology and disorder is richer than previously recognized and highlight that weak disorder can impose practical limits on topological photonic transport. Beyond photonics, the mechanism is broadly relevant to Chern insulators in acoustics, electronics, and magnonics, where "necklace state" defect modes may similarly compromise unidirectional transport. Our work not only deepens the theoretical understanding of the interaction between topological states and disorder, but also provides important guidance for designing the error-free optical communication through topological edge states.

**Methods**

All numerical results presented in this work are simulated using the RF module of COMSOL Multiphysics. The topological characteristics are evaluated through local Chern number calculations and reflection phase winding analyses. Technical details for the former are provided in Section S3 of the Supporting Information, while the specific implementation of the latter is detailed in the second subsection of the Results.

The investigated two-dimensional square lattice photonic crystal is composed of gyromagnetic rods. All gyromagnetic rods have relative permittivity of 13. Each rod can be magnetized individually by permanent magnets attached to the rod ends, which apply a static magnetic field along the $z$ direction.[72] This magnetization causes the rods to exhibit gyromagnetic behavior in the microwave regime, with the relative permeability tensor taking the form $\mu = \begin{pmatrix} \mu_1 & \mu_2 & 0 \\ -\mu_2 & \mu_1 & 0 \\ 0 & 0 & 1 \end{pmatrix}$. Results shown in the work are calculated with $\mu_1 = 1$, and $\mu_2 = 0.4i$, which embodies gyromagnetic properties.[24]

For the configurations used in Figures 1–3, the square array ($30a \times 30a$) has a lattice constant $a = 30$mm. Both the nonmagnetized rods ($A$-type, impurities) and the magnetized rods ($B$-type) have the same radius $r = 0.13a$. For the configuration analyzed in Figure 4, the square array ($30a \times 30a$) has a lattice constant $a = 22.5$mm, with magnetized rods of radius $r_B = 0.13a$ and nonmagnetized rod impurities of radius $r_A = 1.15r_B$. The TM mode dispersion depicted in Figures 1b,4b is simulated using square unit cells with periodic boundary conditions along the $x$ direction and PEC boundary conditions along the $y$ direction. Bulk transmission depicted in Figures 3e,4f is calculated by placing the line source near the center of the left boundary and applying continuity boundary conditions to the top and bottom boundaries, and absorbing conditions to the left and right boundaries.

## ASSOCIATED CONTENT

**Supporting Information**. Statistical analysis confirming effective coupling conditions; histogram of unidirectional transport efficiency for 60 random configurations, full-wave calculation of the local Chern numbers, reflection phase winding in regions above and below the impurity band, and scaling relationship between the system size and the threshold impurity concentration (PDF). This material is available free of charge via the Internet at http://pubs.acs.org.

## AUTHOR INFORMATION

### Corresponding Author

*Lei Zhang – State Key Laboratory of Quantum Optics Technologies and Devices, Institute of Laser Spectroscopy and Collaborative Innovation Center of Extreme Optics, Shanxi University, Taiyuan 030006, China; Email: zhanglei@sxu.edu.cn
*Jun Chen – State Key Laboratory of Quantum Optics Technologies and Devices, Institute of Theoretical Physics and Collaborative Innovation Center of Extreme Optics, Shanxi University, Taiyuan 030006, China; Email: chenjun@sxu.edu.cn

### Author Contributions

†X.S. and T.Q. contributed equally to this work.

### Funding Sources

This work is supported by the National Key R&D Program of China under Grant No. 2022YFA1404003, the National Natural Science Foundation of China (Grants No. 12574340, 12474047, 12174231), the Fund for Shanxi 1331 Project, research project supported by Shanxi Scholarship Council of China.

### Notes

The authors declare no competing financial interest.

## ACKNOWLEDGMENT

The authors gratefully acknowledge helpful discussions with Professors Che-Ting Chan and Zhao-Qing Zhang, Professor Ruo-Yang Zhang, and Postdoctoral Researcher Sen Lin from The Hong Kong University of Science and Technology.